\documentclass[%
 reprint,
 superscriptaddress,
 amsmath,amssymb,
 aps,
 pre,
 showkeys
]{revtex4-1}

\pdfoutput=1

\usepackage{graphicx}
\usepackage{dcolumn}
\usepackage{bm}
\usepackage{hyperref}

\usepackage{mathrsfs}

\usepackage[utf8]{inputenc}
\usepackage[english]{babel} 

\usepackage{fancyhdr}  

\fancypagestyle{general}{
\fancyhf{} 
\lhead{}%
\chead{}%
\rhead{Collective enhancement in dissipative quantum batteries\ \textemdash\ \thepage}
\lfoot{}%
\cfoot{}%
\rfoot{}%
}

\usepackage{tikz}

\providecommand*{\eu}%
{\ensuremath{\mathrm{e}}}
\providecommand*{\iu}%
{\ensuremath{\mathrm{i}}}

\providecommand*{\ped}[1]{%
    \ensuremath{_\mathrm{#1}}}
    
\usepackage{physics}

\usepackage{float}

\usepackage{enumitem}

\addto\captionsenglish{%
  \renewcommand{\figurename}%
    {FIG.}%
}

\begin{document}


\title{Collective enhancement in dissipative quantum batteries}

\author{Javier Carrasco}
 \altaffiliation[Also at ]{Electrical Engineering Department, Faculty of Physical and Mathematical Sciences, University of Chile, Santiago, Chile}
 \email{javier.carrasco@ug.uchile.cl}
\affiliation{%
 Department of Physics, Faculty of Physical and Mathematical Sciences, University of Chile, Santiago, Chile
}%
\author{Jerónimo R. Maze}
\affiliation{%
 Institute of Physics, Pontificia Universidad Católica de Chile, Santiago, Chile
}%
\affiliation{%
 Research Center for Nanotechnology and Advanced Materials, Pontificia Universidad Católica de Chile, Santiago, Chile
}%
\author{Carla Hermann-Avigliano}
\affiliation{%
 Department of Physics, Faculty of Physical and Mathematical Sciences, University of Chile, Santiago, Chile
}%
\affiliation{%
 ANID - Millennium Science Initiative Program - Millennium Institute for Research in Optics (MIRO), Chile
}%
\author{Felipe Barra}
\affiliation{%
 Department of Physics, Faculty of Physical and Mathematical Sciences, University of Chile, Santiago, Chile
}%

\date{\today}


\begin{abstract}
We study a quantum battery made out of $N$ non-mutually interacting qubits coupled to a dissipative single electromagnetic field mode in a resonator. We quantify the charging energy, ergotropy, transfer rate, and power of the system, showing that collective enhancements are still present despite of losses, and can even increase with dissipation. Moreover, we observe that a performance deterioration due to dissipation can be reduced by scaling up the battery size. This is useful for experimental realizations when controlling the quality of the resonator and the number of qubits are limiting factors.
\end{abstract}


\maketitle


\section{Introduction}

The possibility of using quantum resources for technological purposes is currently an active research field, in which quantum batteries (QBs) have emerged as promising tools for the thermodynamic control at the quantum scale \cite{Campaioli2018, Henao2018, Andolina2019_1, Andolina2019_2, Santos2019, Santos2020, Bhattacharjee2021}. A quantum battery is a quantum mechanical system that behaves as an efficient energy storage device. Its realization is motivated by the fact that genuine quantum effects such as entanglement or squeezing can typically boost the performances of classical protocols, e.g., by speeding up the underlying dynamics \cite{Deffner2017, Giovannetti2003}. Enhancements provided by quantum correlations in the charging (or discharging) process of a QB has been previously discussed in \cite{Alicki2013, Hovhannisyan2013, Binder2015, Campaioli2017, Julia-Farre2020, Rossini2020, Rosa2020}. Possible realizable models have been explored, including spin-chains and qubits interacting with electromagnetic fields \cite{Le2018, Ferraro2018, Andolina2018, Zhang2019, Crescente2020, Delmonte2021, Yao2021, Barra2022}. More recently, an experimental realization using organic microcavities has been reported \cite{Quach2022}.

Up to now, research efforts have been mostly focused on understanding QBs as closed systems, isolated from the environment. The dissipation of real QBs has only recently been considered \cite{Farina2019, Barra2019, Pirmoradian2019, Liu2019, Carrega2020, Gherardini2020, Bai2020, Quach2020, Ghosh2021, Santos2021}. Hence, an important question to answer is how dissipation harms the performance of a QB. Understanding collective enhancements in the presence of dissipation is crucial to experimentally realize QBs.

In this Letter we address the aforementioned question by analyzing the case of $N$ non-mutually interacting two-level systems (qubits) charged via a single electromagnetic field mode in a resonator. This system is described by the Tavis-Cummings model \cite{TavisCummings1968, TavisCummings1969}, which is known to provide an effective description of experimentally feasible many-body systems in circuit and cavity QED \cite{Fink2009, Leek2009, Lolli2015, Yang2018}. This configuration is compared to $N$ copies of a resonator with one qubit. The former is a collective QB while the latter is a parallel QB. Our findings show that the performance of parallel and collective QBs (for instance, the power) decreases under dissipation as expected. Nevertheless, the ratio between the power of the collective over the parallel QB increases with dissipation meaning that the deterioration in performance is smaller for the collective QB. More remarkably, we find that the loss in performance due to dissipation can be reduced by scaling up the QB, which means equally increasing the injected energy and number of qubits. In many systems this is easier to do than decreasing dissipation. We also analyze collective enhancements and performances in charging energy, ergotropy, and transfer rate. For the first two, we obtain similar results as for the charging power. For the transfer rate, instead, we find that its collective enhancement decreases and its performance increases as the dissipation rate increases.


\section{Dissipative Quantum Batteries}

\begin{figure}[t]
    \centering
    \includegraphics[width=\linewidth]{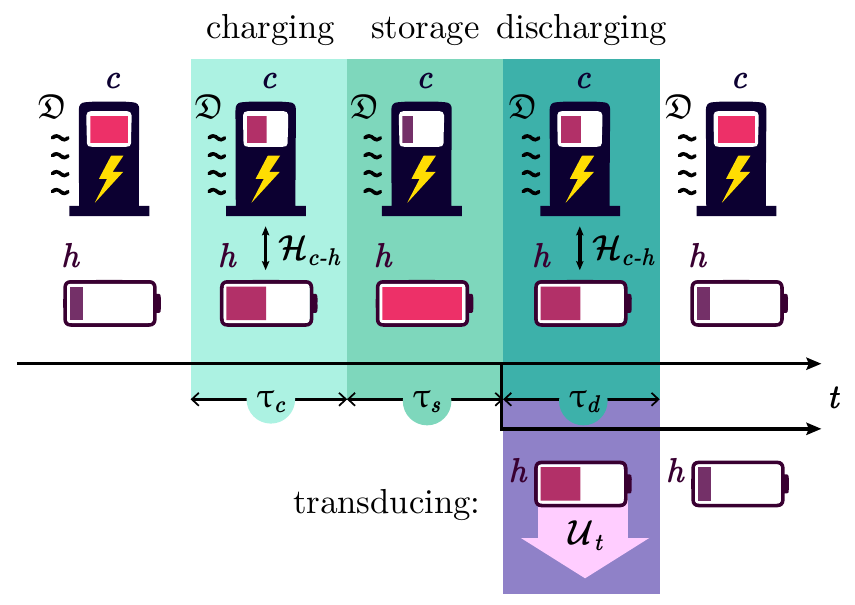}
    \caption{A generic QB consisting of a charger ``$c$'' and a holder ``$h$'' in its three operation stages: charging, storage, and discharging. The holder, initially empty, obtains energy from the charger interacting a time $\tau_c$ with the Hamiltonian $\mathcal{H}_{c-h}$. The energy transferred to the holder is stored for a time $\tau_s$ until we desire to use the energy. To discharge the holder in a time $\tau_d$ we can proceed in transducing mode (bottom) with external classical fields over the holder through a unitary operation $\mathcal{U}_t$, or in normal mode (top) through the charger where the dissipation channels might be used to finally extract the energy. We focus on time scales for which the charger dissipation ($\mathfrak{D}$) is relevant, but holder dissipation is negligible.}
    \label{fig:QB+Stages+Modes}
\end{figure}

We start by defining a general model for a QB. It consists of two subsystems: an energy ``holder'' that stores energy for long times, and an energy ``charger'' that acts as transducer of energy to deliver external input energy into the holder. The charger can also be used as a discharging path for the holder, unless the battery is designed to be directly discharged from the holder via external classical fields. We define the former discharging mode as ``normal'', and the latter as ``transducing''. Furthermore, we identify that the battery can be charged up either from initial conditions or through external classical driving. In this work we focus on the former charging case, where the charger is prepared in an energetic state and then left to interact with the holder allowing for energy transfer. Therefore, the QB system is effectively described by the Hamiltonian
\begin{align}
    \mathcal{H}_{\lambda} = \mathcal{H}_c + \mathcal{H}_h + \lambda\mathcal{H}_{c-h},
\end{align}
where $\mathcal{H}_{c,h}$ are the free Hamiltonians of the charger and holder, respectively, interacting via the Hamiltonian $\mathcal{H}_{c-h}$. The parameter $\lambda$ can be 0 or 1 and defines in which stage the QB is: $\lambda=1$ for the charging and normal discharging stages, and $\lambda=0$ for the storage and transducing discharging stages. For this QB protocol to work, the switching time between stages must be much smaller than the characteristic time of the dynamical evolution of the battery in each separate stage. If this condition is fulfilled, the Hamiltonian $\mathcal{H}_{\lambda}$ is correctly considered time-independent in each separate stage. The previously defined QB model is pictured in Fig. \ref{fig:QB+Stages+Modes}.

We focus on a model in which the charger is a single electromagnetic field mode in a resonator coupled to an array of $N$ non-mutually interacting identical qubits that act as the holder \cite{Ferraro2018}. Under the rotating wave approximation, the Hamiltonian is that of the Tavis-Cummings model \cite{TavisCummings1968, TavisCummings1969, Carmichael2002, Klimov2009, Putz2017, Kirton2019}: $\mathcal{H}_c = \hbar \omega_c a^{\dagger}a, \mathcal{H}_h = \hbar \omega_h \sum_{i=1}^N \sigma_i^{+} \sigma_i^{-}, \mathcal{H}_{c-h} = \hbar g \sum_{i=1}^N (a\sigma_i^{+} + a^{\dagger}\sigma_i^{-})$, where $a$ $(a^{\dagger})$ is a bosonic annihilation (creation) operator, $\sigma_i^{\pm}$ are raising ($+$) or lowering ($-$) atomic operators for the $i$-th qubit, $\omega_{c,h}$ are the characteristic frequencies of the resonator and the qubits, respectively, $g$ is the coupling strength, and $\hbar$ is the reduced Planck's constant. We use $g=10^{-3}\omega_c$ to be consistent with the rotating wave approximation. In this physical system, transitioning between battery stages means to change between negligible ($\lambda=0$) and non-negligible ($\lambda=1$) holder-charger interaction, which can be achieved by external fields that modify $\omega_{c,h}$ to transition between far-from-resonant ($\abs{\omega_c - \omega_h} \ll g,$ hence $\lambda=0$) and resonant ($\omega_c=\omega_h,$ hence $\lambda=1$) dynamics. Hence, $\omega_{c} = \omega_{h}$ during charging and normal discharging stages \cite{Ferraro2018}.

For an isolated QB, the dynamics would simply be determined by the Hamiltonian $\mathcal{H}_{\lambda}$. However, to account for dissipation and decoherence on the QB due to its coupling with the environment, a master equation is needed where a dissipator super-operator $\mathfrak{D}[\cdot]$ is added to the Liouville-von Neumann equation giving \cite{Breuer2002, Rivas2011, Carmichael2002}
\begin{align}
    \dv{\rho}{t} = -\frac{\iu}{\hbar} \comm{\mathcal{H}_{\lambda}}{\rho} + \mathfrak{D}[\rho],
\end{align}
where $\rho$ is the density operator or state of the QB. Moreover, we argue that a QB has practical sense only if the dissipation rate of the holder, $\kappa_h$, is much lower than that of the charger, $\kappa_c$. Hence, if the frequency holder-charger effective coupling factor is $g\ped{eff} \gg \kappa_h$ but comparable to $\kappa_c$, it is sufficient to consider only the charger dissipator $\mathfrak{D}_c$ during charging and discharging stages. Thus we set $\mathfrak{D} \approx \mathfrak{D}_c$ in the master equation. During the storage stage, holder dissipators might be added. However, storage times $\tau_s \ll 2\pi/\kappa_h$ where holder dissipation is negligible are still relevant as a battery if $\tau_s \gg 2\pi/\kappa_c$, meaning that we could achieve energy storage beyond what would be possible with just the charger.

For our Tavis-Cummings QB in study, $g\ped{eff}= g\sqrt{N}$ \cite{Haroche2006, Putz2017} and the charger dissipator is $\mathfrak{D}_c [\cdot] = \kappa (n\ped{th} + 1) (2a\cdot a^{\dagger} - a^{\dagger} a \cdot - \cdot a^{\dagger} a) + \kappa n\ped{th} (2 a^{\dagger} \cdot a - a a^{\dagger} \cdot - \cdot a a^{\dagger})$, which is given by an environment of thermal photons with mean number $n\ped{th} = 1/(\eu^{\hbar\omega_c/k\ped{B}T} - 1)$ at temperature $T$, $2\kappa$ is the charger dissipation rate $\kappa_c$, and $k\ped{B}$ is Boltzmann's constant. We use $n\ped{th}=0$ to solely study the dissipation from the charger avoiding the re-population effects of a nonzero temperature.

\section{Collective Enhancement}

\begin{figure}[t]
    \centering
    \includegraphics[width=\linewidth]{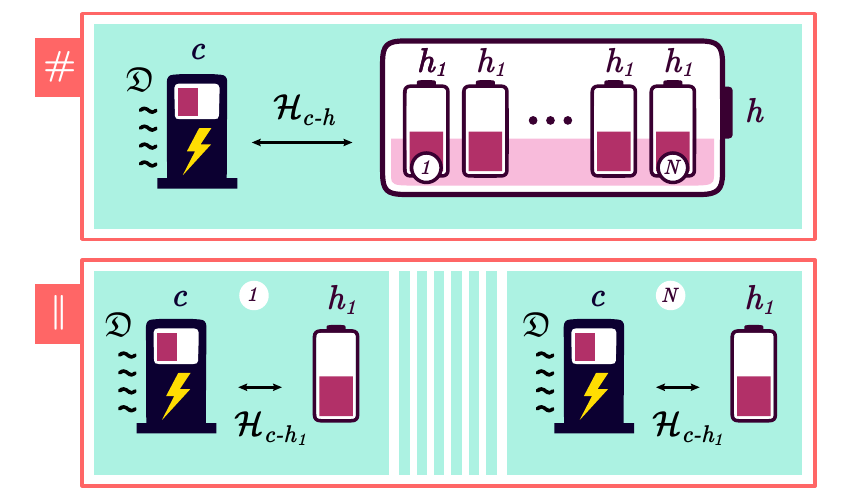}
    \caption{Collective (\#) and parallel ($\parallel$) versions of a QB of size $N$. The collective QB has one charger ``$c$'' with $E_c$ initial energy and one holder ``$h$'' consisting in $N$ units cells ``$h_1$''. The parallel QB has $N$ copies of ``$c$'' with $E_c/N$ initial energy, each interacting with a different unit cell ``$h_1$'' as holder. Hence, the $\parallel$ QB is just $N$ copies of the $N$-times down-scaled \# QB.}
    \label{fig:collective_vs_parallel}
\end{figure}

If the holder of a QB consists in $N$ copies of a subsystem, i.e. $N$ holder unit cells, we can compare its performance versus $N$ copies of the QB with only one holder unit cell. These represent two versions of a QB that have the same maximum storable energy. The former is the collective version ($\#$) as it has $N$ holder unit cells interacting with the same charger, while the latter is the parallel version ($\parallel$) as it has $N$ independent copies of one charger with one holder unit cell (see Fig. \ref{fig:collective_vs_parallel}). When compared, the collective version shows an increase in the transfer rate due to correlations \cite{Alicki2013, Hovhannisyan2013, Binder2015, Campaioli2017}. This motivates the definition of the transfer rate collective enhancement (similar to the quantum advantage \cite{Campaioli2018}) $\Gamma_{1/\bar{\tau}} = (1/\bar{\tau}^{\#})/(1/\bar{\tau}^{\parallel}) \equiv \bar{\Omega}^{\#} / \bar{\Omega}^{\parallel}$ where $\bar{\tau}^{\#}$ and $\bar{\tau}^{\parallel}$ are the interaction times needed to charge or discharge the QB up to its first dynamical maximum in the collective and parallel versions, respectively. $\Gamma_{1/\bar{\tau}}$ quantifies how much larger is the transfer rate $\bar{\Omega} \equiv 2\pi/\bar{\tau}$ for the collective version compared to the parallel one. If $\Gamma_{1/\bar{\tau}} > 1$, the collective QB has a larger transfer rate than the parallel QB, meaning that $\bar{\Omega}$ is enhanced in the collective version of the QB. We extend the previous concept to all interesting figures of merit $f$ of a QB that quantify its performance (larger $f$, better battery), including charging and discharging energy, transfer rate, power, and maximum extractable energy through unitary operations (ergotropy \cite{Allahverdyan2004}), by defining the collective enhancements $\Gamma_{\bar{f}} = \bar{f}^{\#}/\bar{f}^{\parallel}$. More generally, if $\Gamma_{\bar{f}}>1$, $f$ is enhanced in the collective version. For isolated QBs, $\Gamma_{1/\bar{\tau}}>1$ and $\Gamma_{\bar{E}_h}<1$, where $E_h$ is the holder energy, and the bar represents the event of first maximum transferred energy into the holder during the charging stage (see Fig. \ref{fig:exampleRabiOsc+MaxPoint}). However, in the thermodynamic limit ($N \rightarrow \infty$) these inequalities might change \cite{Julia-Farre2020}. In general, the figures of merit depend on the duration $\tau$ of the charging ($\tau_c$), storage ($\tau_s$) or discharging ($\tau_d$) stages depicted in Fig. \ref{fig:QB+Stages+Modes}, and $\bar{f}=f(\bar{\tau})$. In this work we show that the former inequality also holds for non-isolated QBs, while the latter changes.


\section{Results}

\begin{figure}[htb]
    \centering
    \includegraphics[width=.95\linewidth]{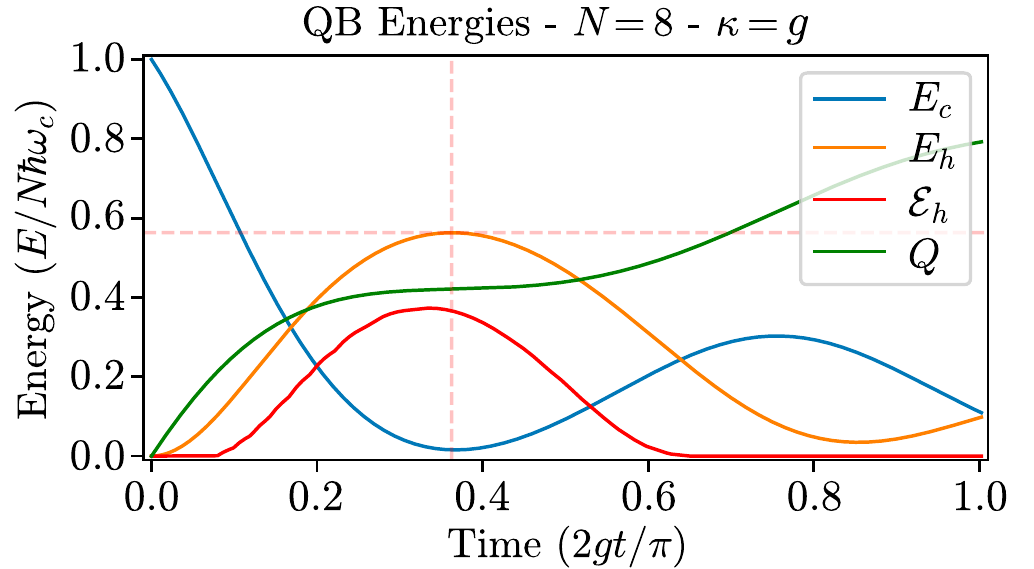}
    \caption{Mean energy $E_h$ and ergotropy $\mathcal{E}_h$ charged on the holder (array of $N$ qubits) when charging from an initial Fock light state. The mean energy $E_c$ of the charger (resonator) and heat $Q$ from the QB are also shown. The dynamics seen correspond to \emph{Rabi-like} oscillations, transferring energy from the charger to the holder. The holder oscillations for different $N$ and $\kappa$ are given in Appendix \ref{ap:AppendixC}. The segmented lines define $\bar{\tau}$ and $\bar{E}_h$.}
    \label{fig:exampleRabiOsc+MaxPoint}
\end{figure}

\begin{figure}[htb]
    \centering
    \includegraphics[width=\linewidth, keepaspectratio]{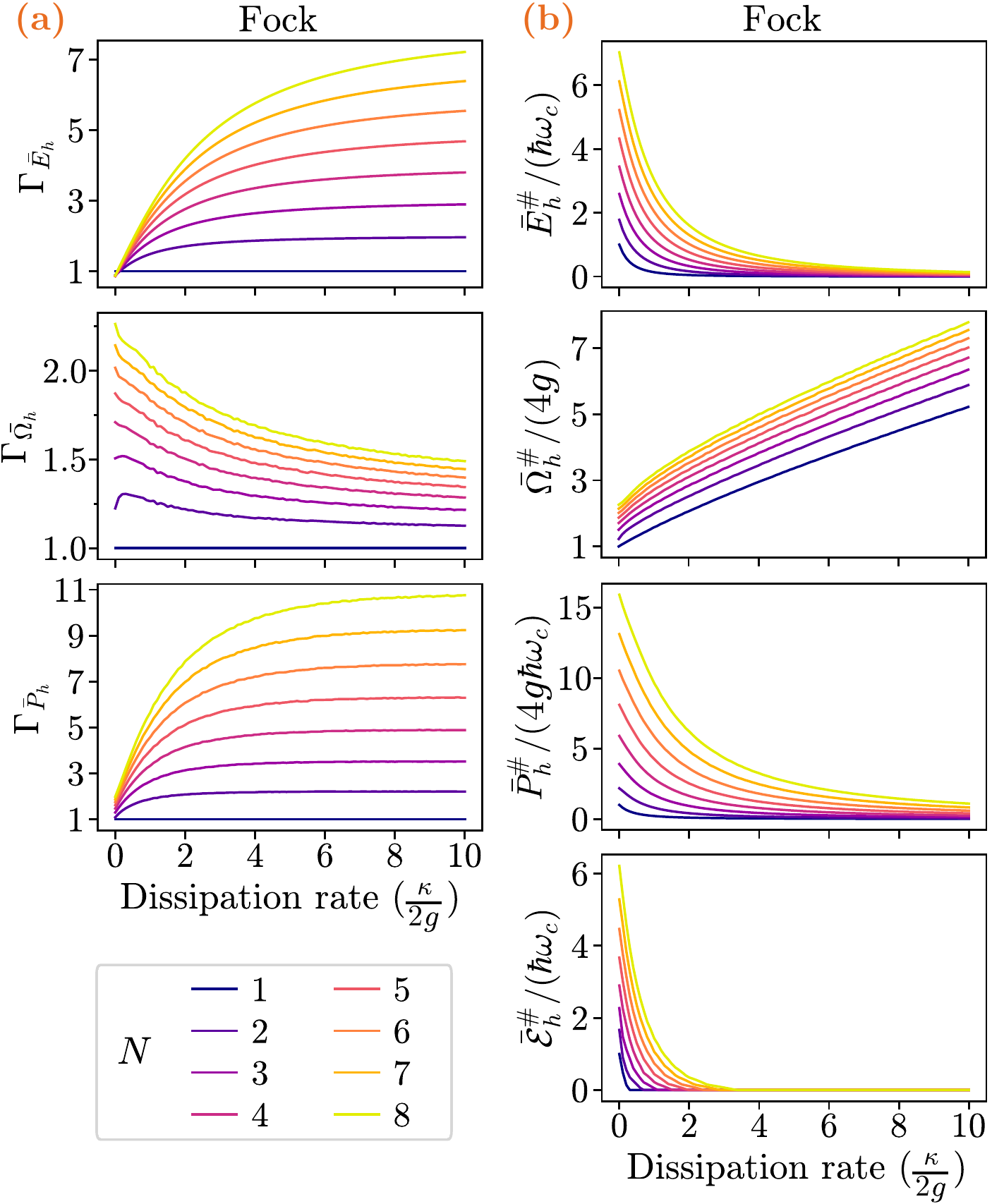}
    \caption{Collective (a) enhancements $\Gamma_{\bar{f}}$ and (b) performances for the first maximum charge when charging from Fock initial conditions for energy, transfer rate, power, and ergotropy. The results for coherent and thermal initial conditions are given in Appendix \ref{ap:AppendixC}.}
    \label{fig:result1}
\end{figure}

\begin{figure}[htb]
    \centering
    \includegraphics[width=\linewidth, keepaspectratio]{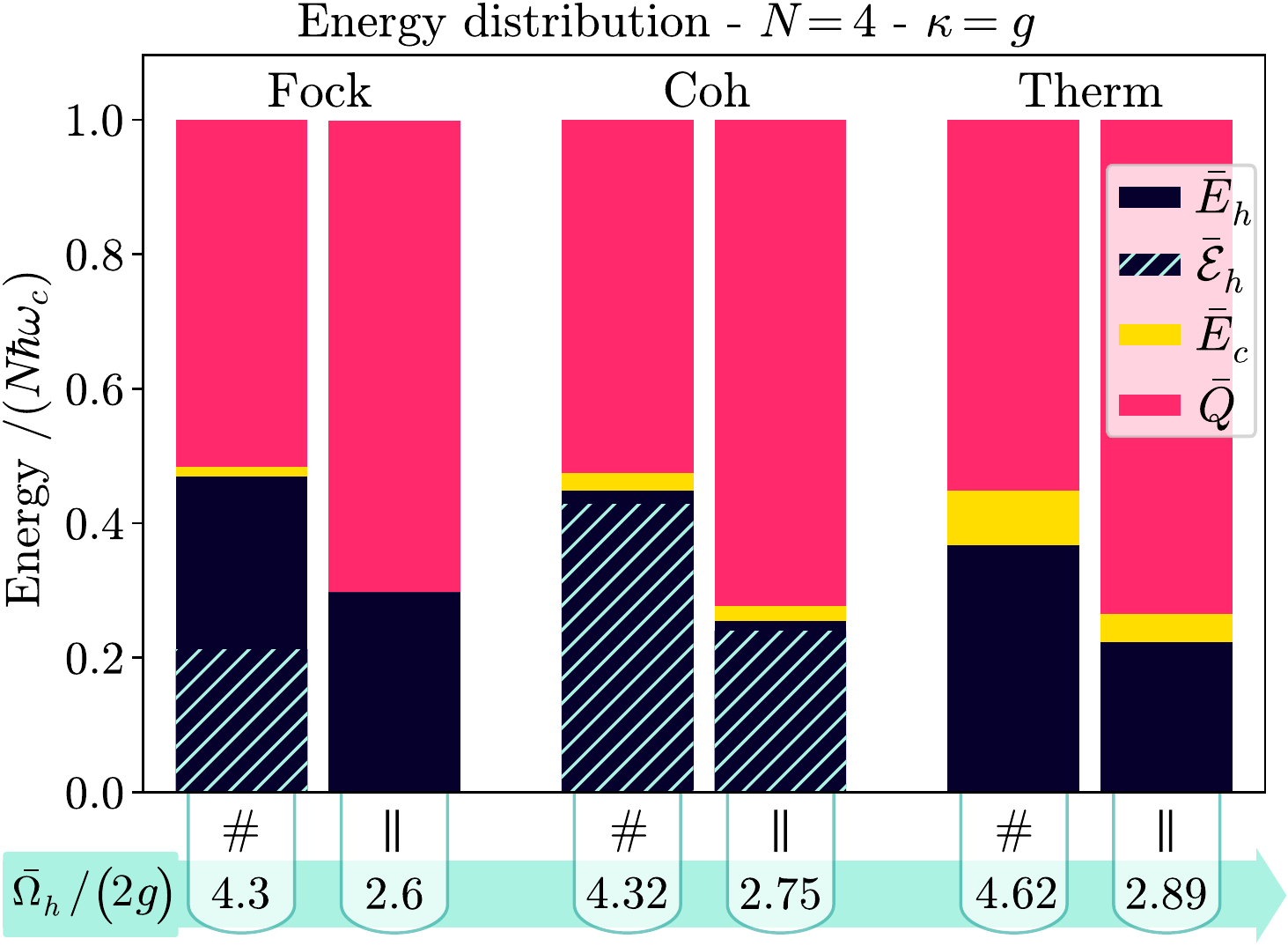}
    \caption{Distribution of the total energy over the QB. Collective (\#) and parallel ($\parallel$) are compared for the three cases of initial conditions simulated.}
    \label{fig:result2}
\end{figure}

We now study the robustness of the collective enhancements when dissipation is taken into account in the Tavis-Cummings QB. The following initial conditions of the resonator are studied: Fock, coherent, and thermal states. All three with mean number of photons $\Tr\{a^{\dagger}a \Tr_h\{\rho(0)\}\} = N$. In all cases, the qubits are initially in their ground state. We analyze the holder (charger) energy $E_{h(c)}$, transfer rate $\Omega_{h(c)} \equiv 2\pi/\tau$, power $P_{h(c)} \equiv E_{h(c)} / \tau$, and ergotropy $\mathcal{E}_{h(c)}$ for the charging stage. Fig. \ref{fig:exampleRabiOsc+MaxPoint} depicts $E_{h}$, $E_{c}$,  the holder ergotropy, which is computed from the reduced state of the holder at the given time, and the heat from the QB, $Q(t) \equiv -\int_0^t \Tr\{\mathcal{H}(t')\mathfrak{D}[\rho(t')]\} \dd t'$, over time. 
The total energy is given by $E_h(t) + E_c(t) + Q(t) = N\hbar\omega_c$. Frequencies are shown in terms of the Rabi frequency of the Fock case when $N=1$ and $\kappa=0$, given by $\Omega_R \equiv 2g$. Fig. \ref{fig:result1} shows the collective enhancements $\Gamma_{\bar{f}}$ and performances $\bar{f}$ for the Fock case for different values of $\kappa$ and $N$. The other two cases produce very similar results and hence are included in Appendix \ref{ap:AppendixC}. Nonetheless, the key differences between them are visualized in Fig. \ref{fig:result2} for the respective collective and parallel QBs.

Naturally, larger dissipation rates produce faster loss of energy available to charge up the holder resulting in a decrease of $\bar{E}_h^{\#}$ as well as $\bar{\mathcal{E}}_h^{\#}$ (see Fig. \ref{fig:result1}). On the other hand, larger $\kappa$ also means that the first maximum transferred energy into the holder, although smaller, is reached sooner, and therefore the transfer rate performance increases (see Fig. \ref{fig:result1}). Hence, as the decrease of $\bar{E}_h^{\#}$ shows to be higher than the increase of $\bar{\Omega}_h^{\#}$, the charging power $\bar{P}_h^{\#}$ ends up decreasing. These results are plotted for different $N$ in Fig. \ref{fig:result1}.b, which shows that the curves move upwards when $N$ increases, meaning that a reduction in performance due to dissipation may be restored by scaling up the QB. For $\kappa=0$ this behavior was obtained in \cite{Ferraro2018}, and our results for $\kappa>0$ can here be considered as an extrapolation of those results. Although, caution must be taken with the statement about restoring the performance, because the separation between curves of consecutive values of $N$ seems to get smaller in each increment for a fixed $\kappa$, and even more as $\kappa$ is also increased. However, for values of $\kappa$ around $4g$ or smaller, we argue that we do not need to fabricate a resonator with the highest possible quality factor as long as we compensate by scaling up the QB. This can be of practical advantage for realizing a QB with systems where adding more qubits and photonic energy is feasible, as in the case of NV centers, which come as natural defects in diamond crystals \cite{Schirhagl2014}.

The increase of performances with $N$ is explained (for $\kappa = 0$ \cite{Klimov2009}) by the collective behavior of the qubits producing a larger effective Rabi frequency, allowing for a faster population of the excited eigenenergies. This upgrade of the performances is better quantified as collective enhancement in Fig. \ref{fig:result1}.a, where it is clear that the collective QB version is better than the parallel one ($\Gamma_{\bar{f}}>1$), except for the energies at $\kappa \approx 0$ where $\Gamma_{\bar{E}_h}\leq 1$. This exception is important because it shows that a simple extrapolation of the collective enhancement in energy obtained at $\kappa=0$ would give the wrong result as if the parallel battery were to perform better. Moreover, we observe that collective enhancements in energy and power increase for larger dissipation rates, which means that despite of the decrease of the respective performances in both QB versions, the collective ones decrease less, i.e. they are more robust under dissipation. Instead, the collective enhancement in charging transfer rate decreases with larger $\kappa$, meaning that the parallel QB is more robust in this regard. 

Regarding the ergotropy, its collective enhancement is not shown as the numerator and denominator become (non-simultaneously) exactly zero very often. Except for values of $\kappa \sim 0$, we get $\Gamma_{\bar{\mathcal{E}}} > 1$, meaning that for transducing QBs more energy is available in the collective version. This is shown in Fig. \ref{fig:result2}, specially clear for the Fock case. Fig. \ref{fig:result2} shows the energy distributed in its different available forms, and from here it is concluded that the coherent case is the best. This happens because the coherent case is much more robust than the Fock one against dissipation. Only for $\kappa \approx 0$ and $N \approx 1$ the Fock case is better as seen in Appendix \ref{ap:AppendixC}. For $\kappa=0$ and large enough $N$, the coherent case is better because when the initial condition is a coherent state, the amount of energy locked in the charger-holder correlations is minimized as studied in \cite{Andolina2019_2}.

In the discharging stage, the initial condition of the holder is the one reached after the charging process, and the charger is empty. The subsequent dynamics is again a Rabi-like oscillation, as shown in Fig. \ref{fig:exampleRabiOsc+MaxPoint}. Therefore, similar results of performances and collective enhancements are expected for the normal discharging stage.


\section{Conclusions}

In this Letter we have defined different types of open QBs and analyzed them in the context of the Tavis-Cummings model, studying their collective behavior under dissipation, considering three types of initial conditions. For this study, we have defined the collective enhancements as figures of merit for the performance of a collective QB over its parallel version. In particular, we have also focused on the case of a QB made of a photonic resonator and an array of qubits to obtain quantitative estimations.

We observed that dissipation, while diminishing the performance in charging energy and power, the corresponding collective enhancements are instead boosted up, because the collective QB version is more robust against dissipation. Regarding the charging transfer rate, the opposite occurs as the parallel version is more robust in this quantity. In addition, we showed that, a higher collective performance can be achieved by scaling up the QB, i.e. increasing the number of qubits and photonic energy in the same amount, although due to the decrease in energy performance, this effect is only meaningful for $\kappa \lesssim 4g$. Furthermore, this up-scaling increases all collective enhancements. Also, we showed that initially preparing the resonator light mode in a coherent state produces better results in ergotropy performance than a Fock or thermal state.

Our work shows that collective QBs are better than parallel ones in a different way than simple extrapolation from results of isolated QBs ($\kappa=0$) as we commented for the energy collective enhancement. It also shows that the dissipation rate $\kappa_c$ and the scale $N$ of the QB can be seen as degrees of freedom to design a QB with a desired performance, while understanding that collective QBs are more robust under changes in dissipation. This is particularly useful if increasing $N$ is much easier than decreasing $\kappa_c$ in the specific system to experimentally realize, or if the precision error in obtaining a specific $\kappa_c$ is important.

Finally, the collective enhancements and performances introduced here, along the clarification of different types of QBs, will serve for future studies in the field, considering different batteries and their experimental realizations. 

All simulations were performed with QuTiP \cite{QuTiP2} in Python.



\section{Acknowledgments}
J.R.M. acknowledges support from Conicyt Fondecyt Regular grant No. 1180673, Conicyt PIA/Anillo ACT192023, AFOSR grant No. FA9550-18-1-0513, and ONR grant No. N62909-18-1-2180. C.H-A. acknowledges support from Fondecyt Grant Nº 11190078, and Conicyt-PAI grant  77180003, and ANID - Millennium Science Initiative Program - ICN17\_012. J.C. and F.B. acknowledge support from Fondecyt grant No. 1191441, and the ANID Millennium Nucleus ``Physics of active matter''.


\appendix

\section{\label{ap:AppendixA}Ergotropy of the Jaynes-Cummings Model}

If the state $\rho$ and (non-degenerate) Hamiltonian $\mathcal{H}$ of a system are written orderly as
\begin{align}
    \label{eq:ergotropy_formula}
    \rho &= \sum_k r_k \dyad{r_k}, \ r_k \geq r_{k+1},\\
    \mathcal{H} &= \sum_k \epsilon_k \dyad{\epsilon_k}, \ \epsilon_k < \epsilon_{k+1},
\end{align}
where $r_k, \epsilon_k$ are the eigenvalues and $\ket{r_k}, \ket{\epsilon_k}$ the eigenkets of the state and Hamiltonian, respectively; then the ergotropy of $\rho$ can be calculated by \cite{Allahverdyan2004}
\begin{align}
    \mathcal{E}[\rho] = \sum_{jk} r_{j} \epsilon_k (\abs{\braket{r_j}{\epsilon_k}}^2 - \delta_{jk}),
\end{align}
where $\delta_{jk}$ is the Kronecker delta. Thus, for the case of the Jaynes-Cummings model ($N=1$) with a Fock state as the charger initial condition, focusing on the qubit sub-system (tracing over the resonator), it is easy to obtain the state of the qubit
\begin{align}
    \rho = \sin^2(gt)\dyad{e} + \cos^2(gt)\dyad{g},
\end{align}
which is diagonal in the energy basis with eigenvalues $\lambda_1(t) = \sin^2(gt)$ and $\lambda_2(t) = \cos^2(gt)$ corresponding to the eigenkets $\ket{e}$ and $\ket{g}$, respectively.

It can be seen that if
\begin{align}
    \lambda_1(t) \geq \lambda_2(t) &\Longleftrightarrow \cos(2gt) \leq 0\\
    &\Longleftrightarrow gt \in \left[ \frac{\pi}{4}, \frac{3\pi}{4} \right] \pm k\pi, k\in \mathbb{N}.
\end{align}
Therefore, outside of these intervals, the ordering of the eigenvalues is inverted.

Now, in units of $\hbar\omega_c$, $H=\sigma_z/2$ with eigenvalues $\epsilon_1 = -1/2$, $\epsilon_2 = 1/2$ and eigenkets $\ket{\epsilon_1} = \ket{g}$, $\ket{\epsilon_2} = \ket{e}$.

Finally, by simply applying the formula \ref{eq:ergotropy_formula}, the ergotropy is obtained as
\begin{align}
    \mathcal{E}(t) =
    \begin{cases}
    1 - 2\cos^2(gt), \ &gt \in \left[ \frac{\pi}{4}, \frac{3\pi}{4} \right] \pm k\pi \\
    \qq{} 0 \ &gt \not\in \left[ \frac{\pi}{4}, \frac{3\pi}{4} \right] \pm k\pi
    \end{cases},
\end{align}
which corresponds to $\mathcal{E}_h / (\hbar \omega_c)$, and clearly has the first maximum at $t = 2\pi/4g \equiv \pi/\Omega_R$ (used as timescale for all plots) with a value $\bar{\mathcal{E}}_h = \hbar\omega_c$, after recovering the original units in energy.

For the Tavis-Cummings model ($N \geq 1$), the equivalent numerical procedure is done to calculate the ergotropy.

\section{\label{ap:AppendixB}Typical Experimental Values}

For our Tavis-Cummings QB in study, $g\ped{eff}= g\sqrt{N}$ \cite{Haroche2006, Putz2017} and typical experimental values are $\kappa_h \approx 0.5 \times 10^{-2} g\ped{eff}$, with $g\ped{eff} \approx 0.5 \times 10^{-3} \omega_c$, achieved with NV centers as qubits \cite{Doherty2013, Putz2017, Eisenach2021}. In the case of using Rydberg atoms as qubits, the typical values are $\kappa_h \approx 10^{-2} g$, with $g \approx \times 10^{-6} \omega_c$ \cite{Haroche2006, Guerlin2007, Uria2020}. Regarding $\kappa_c$, its value depends on the quality factor of the fabricated resonator, which can reach values such that the charger dissipation can be neglected. However, we study the case in which this is not a precondition.

\section{\label{ap:AppendixC}Additional Simulations Data}

\begin{figure*}[htb]
    \centering
    \includegraphics[scale=0.86]{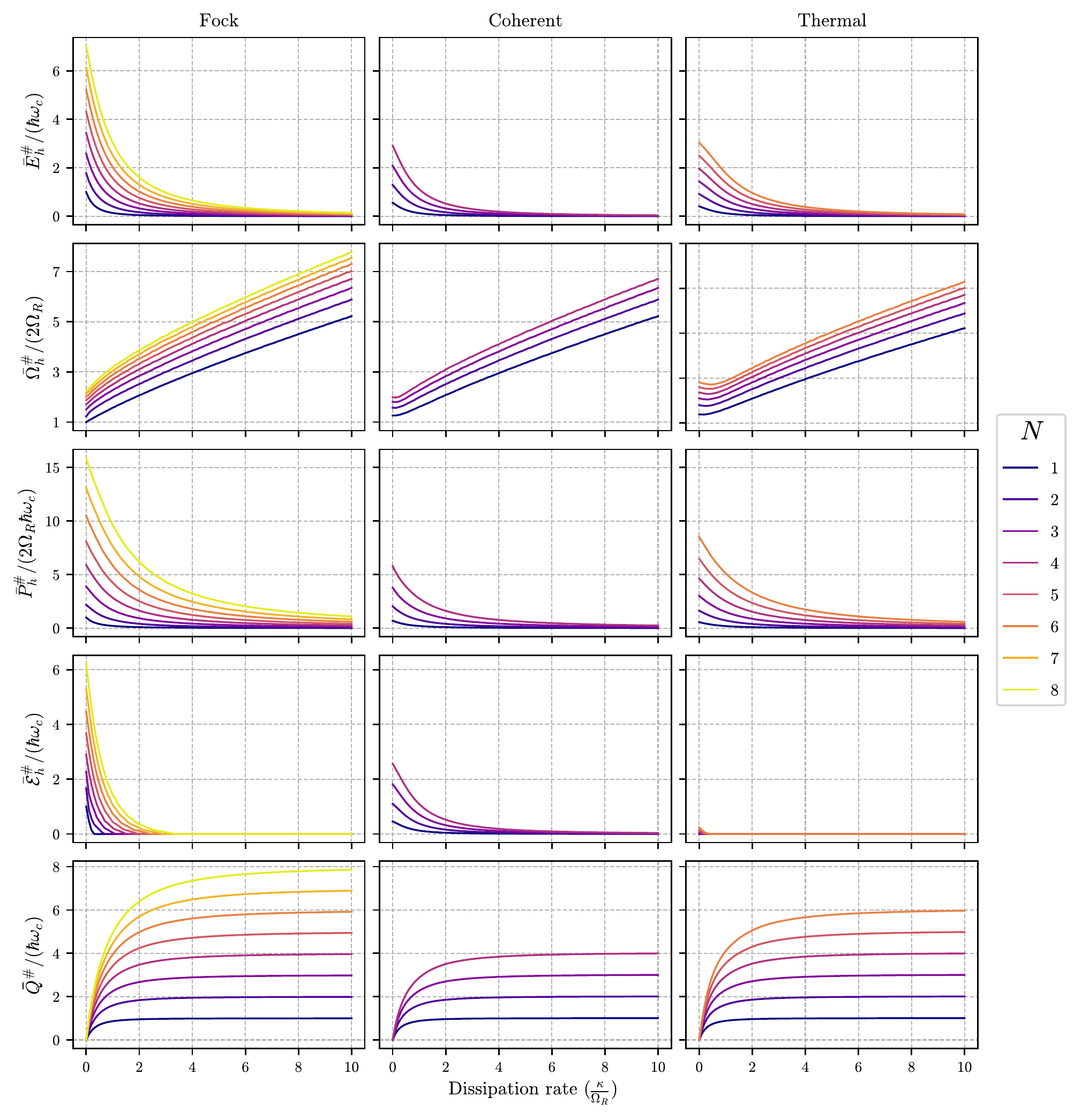}
    \caption{Performances of the collective QB for Fock, coherent and thermal states as the charger initial condition. Comparing between these cases of initial condition, there is no significant meaningful difference, justifying the use of only the Fock case in the main manuscript.}
    \label{fig:SM_fig1}
\end{figure*}

\begin{figure*}[htb]
    \centering
    \includegraphics[scale=0.86]{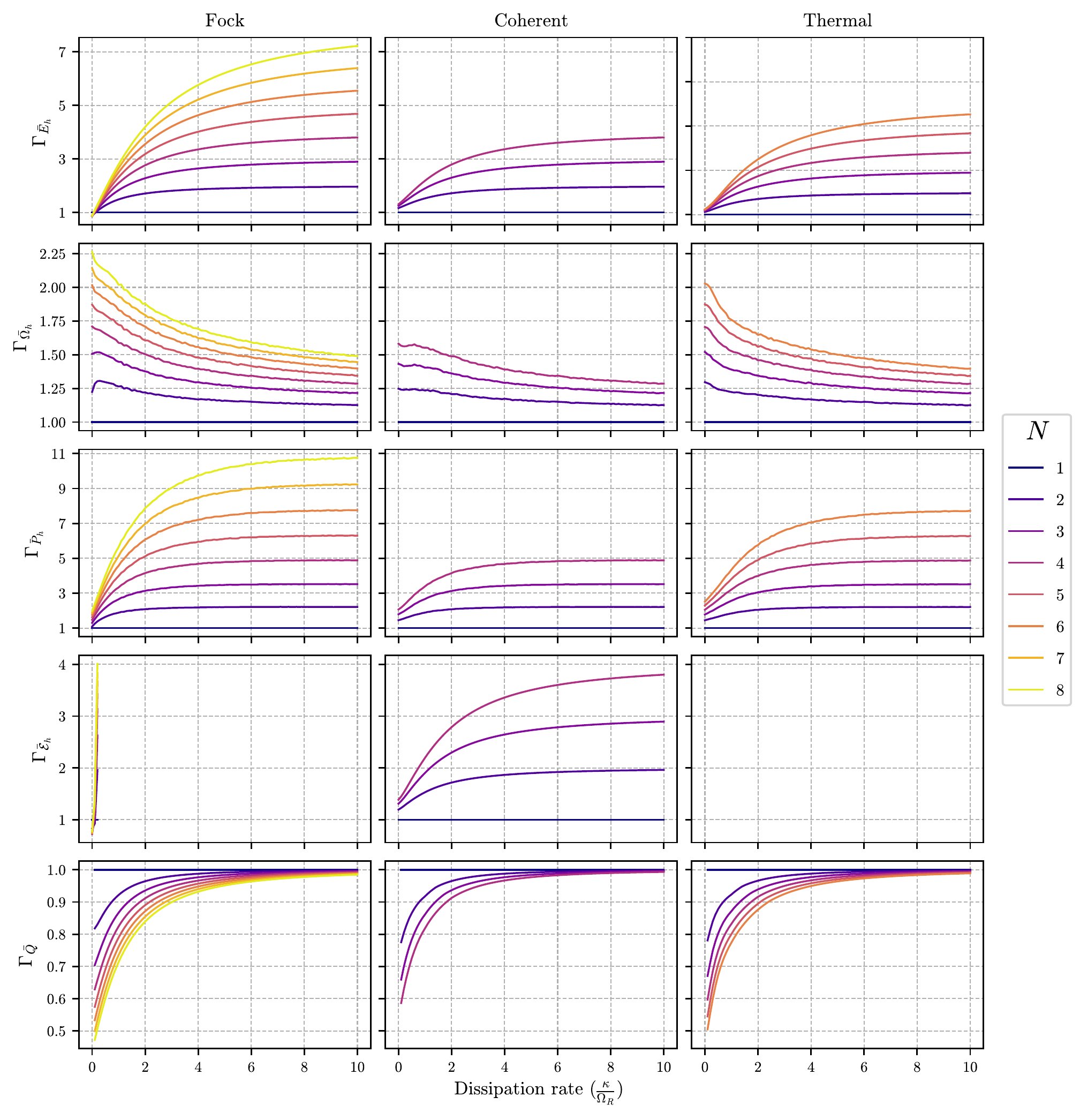}
    \caption{Collective enhancements of the relevant figures of merit for Fock, coherent and thermal states as the charger initial condition. Comparing between these cases of initial condition, there is no significant meaningful difference, excepting the ergotropy, justifying the use of only the Fock case in the main manuscript, where the ergotropy is analyzed separately.}
    \label{fig:SM_fig3}
\end{figure*}

Simulation results complementary to Fig. \ref{fig:result1} of the main text are shown in Figs. \ref{fig:SM_fig1} to \ref{fig:SM_fig3}, where the cases of coherent and thermal states as the charger initial condition are included, as well as the heat from the QB. The coherent and thermal cases have been simulated up to $N=4$ and $N=6$, respectively, due to the rapid increment in numerical complexity.

\begin{figure*}[htb]
    \centering
    \includegraphics[width=\linewidth, keepaspectratio]{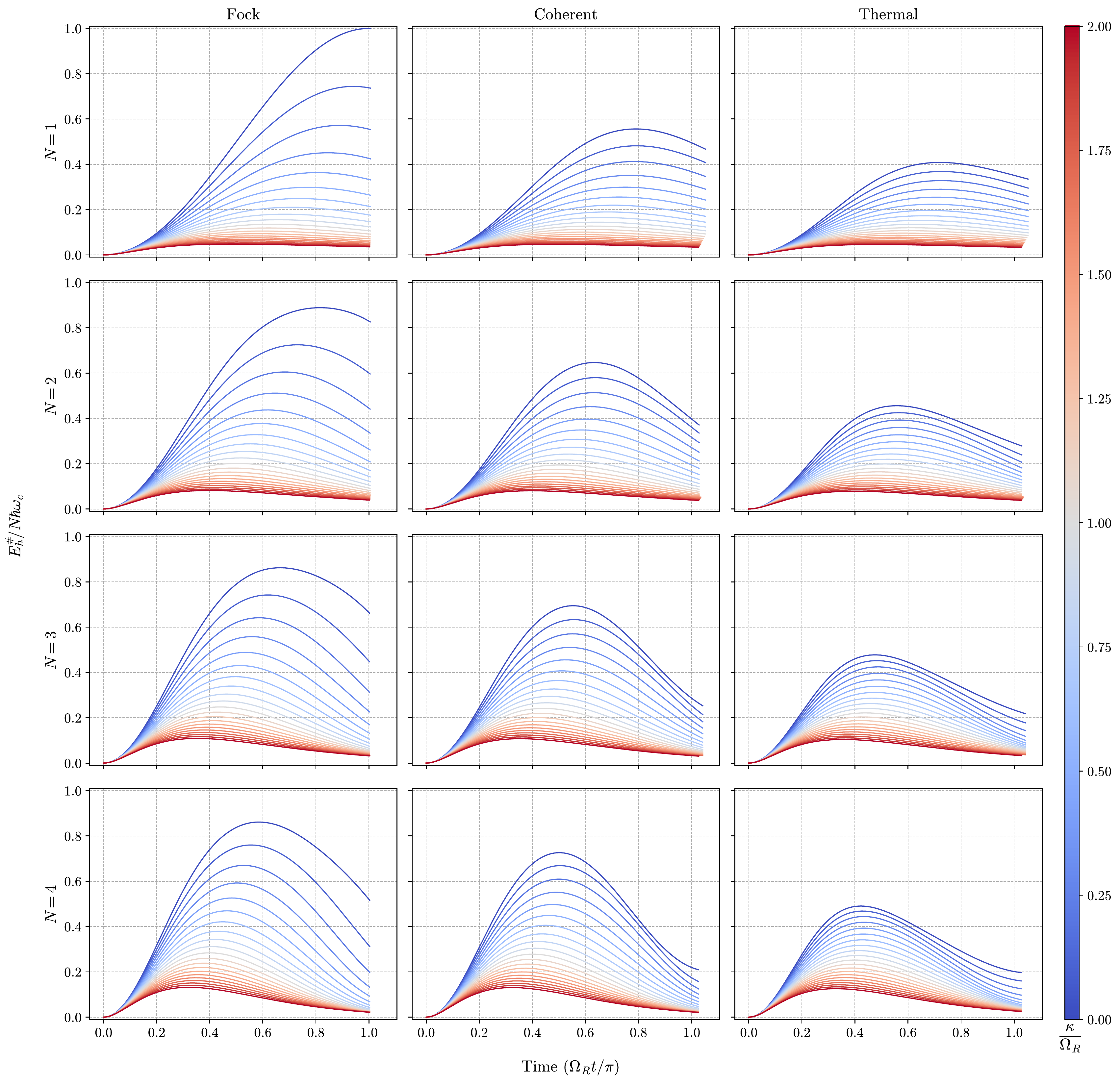}
    \caption{Holder energy over time up to near the first maximum for Fock, coherent and thermal states as the charger initial condition. Results of the collective QB version. It it observed that the Fock case reaches a higher first maximum of holder energy $E_h$ than the coherent and thermal cases. Nonetheless, these maximums are obtained in a shorter time in the thermal case. Also, as the dissipation rate growths, the three cases tend to the same curve.}
    \label{fig:SM_fig5}
\end{figure*}

\begin{figure*}[htb]
    \centering
    \includegraphics[width=\linewidth, keepaspectratio]{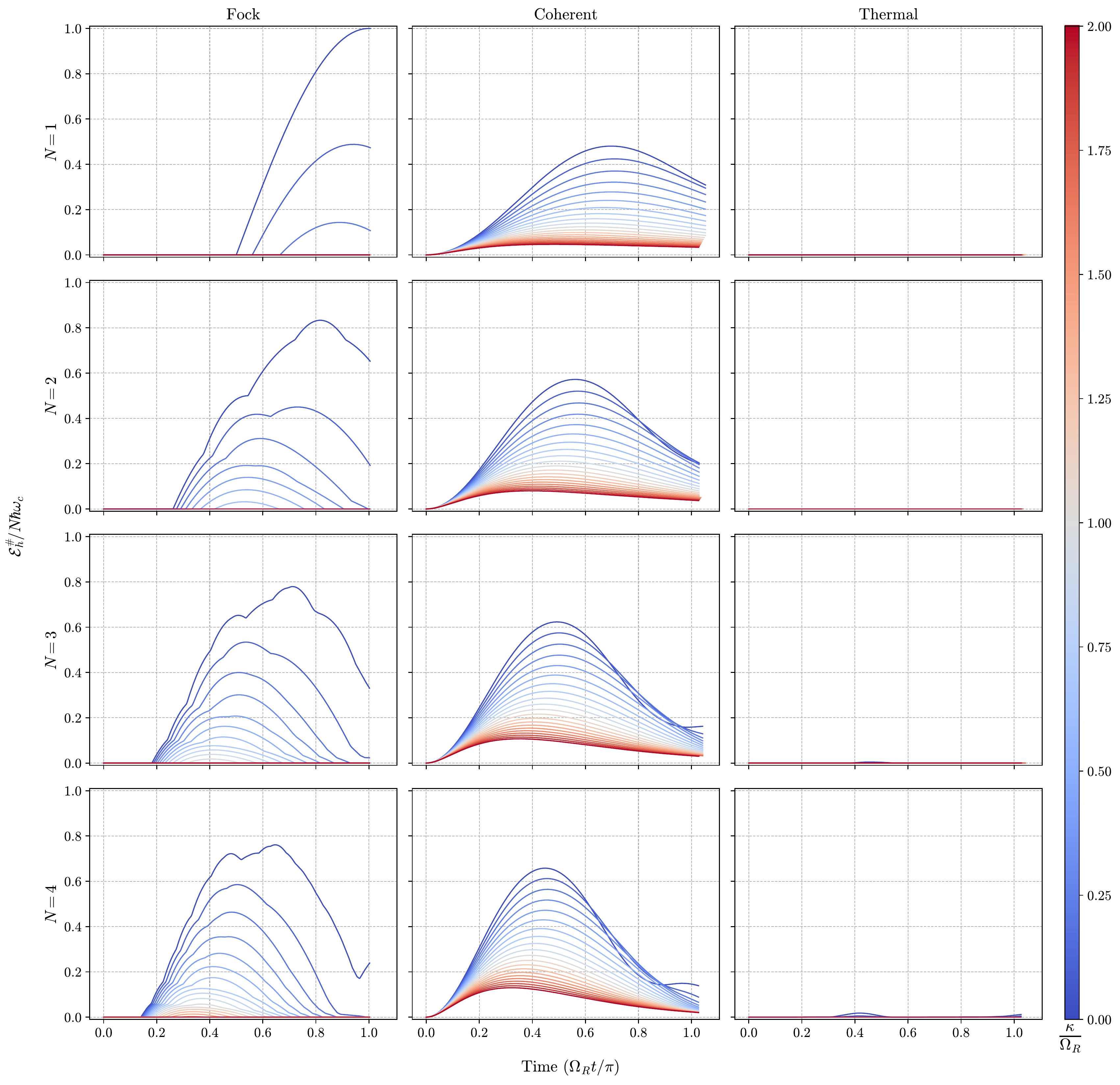}
    \caption{Holder ergotropy over time up to near the first holder energy maximum for Fock, coherent and thermal states as the charger initial condition. Results of the collective QB version. The Fock case shows the highest ergotropy for $\kappa \approx 0$, but it rapidly goes to zero as $\kappa$ increases. Instead, the coherent case changes less, which means that its ergotropy performance is more robust under dissipation.}
    \label{fig:SM_fig6}
\end{figure*}

\begin{figure*}[htb]
    \centering
    \includegraphics[width=\linewidth, keepaspectratio]{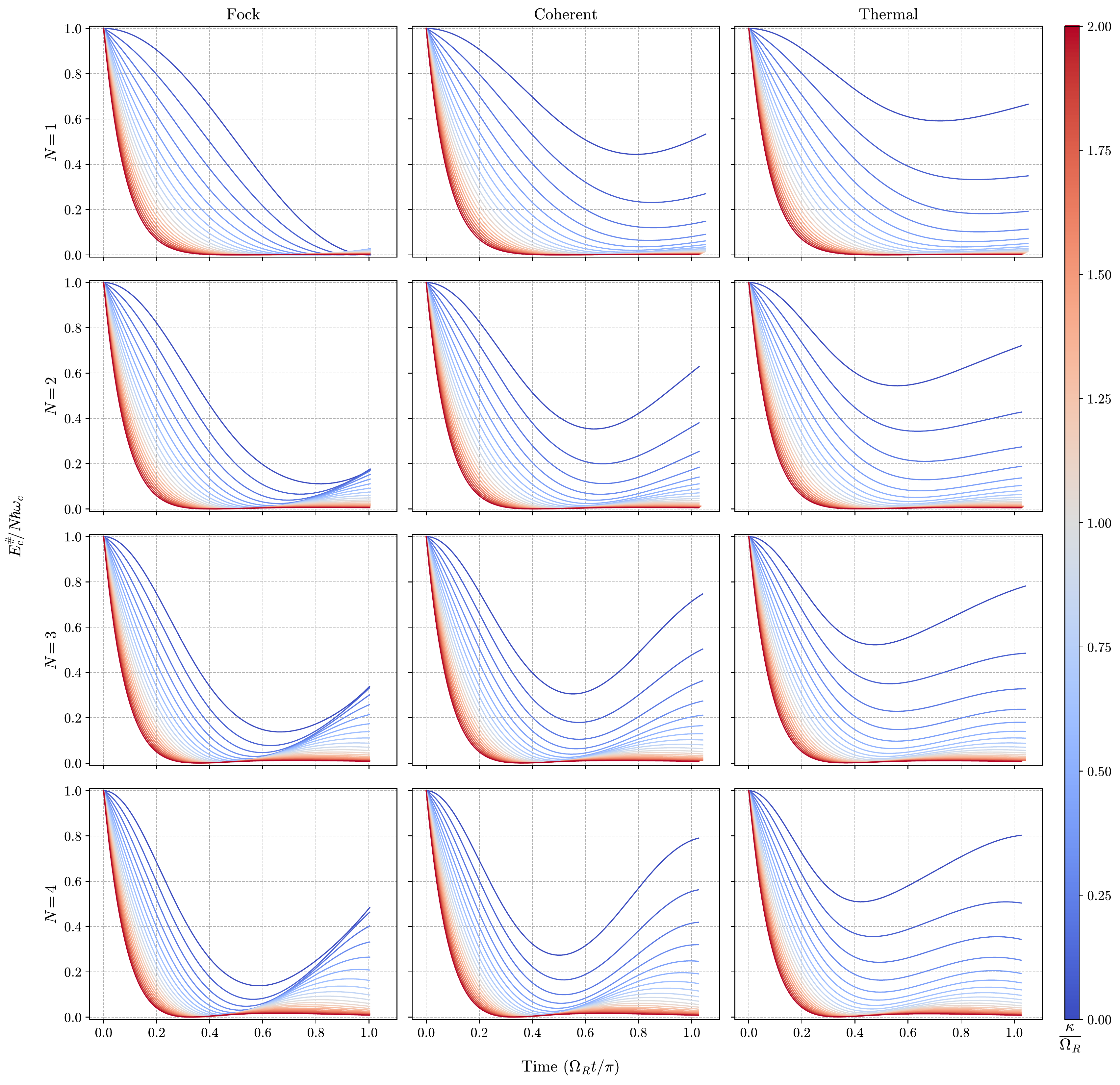}
    \caption{Resonator energy over time up to near the first holder energy maximum for Fock, coherent and thermal states as the charger initial condition. Results of the collective QB version. These plots can be seen as the complement of Fig. \ref{fig:SM_fig5}, since it shows how the resonator energy $E_r$ is reduced while it is been transferred to the holder as $E_h$. Although, $E_r + E_h$ is not a constant because some energy is lost in form of heat $Q$, shown in Fig. \ref{fig:SM_fig8}.}
    \label{fig:SM_fig7}
\end{figure*}

\begin{figure*}[htb]
    \centering
    \includegraphics[width=\linewidth, keepaspectratio]{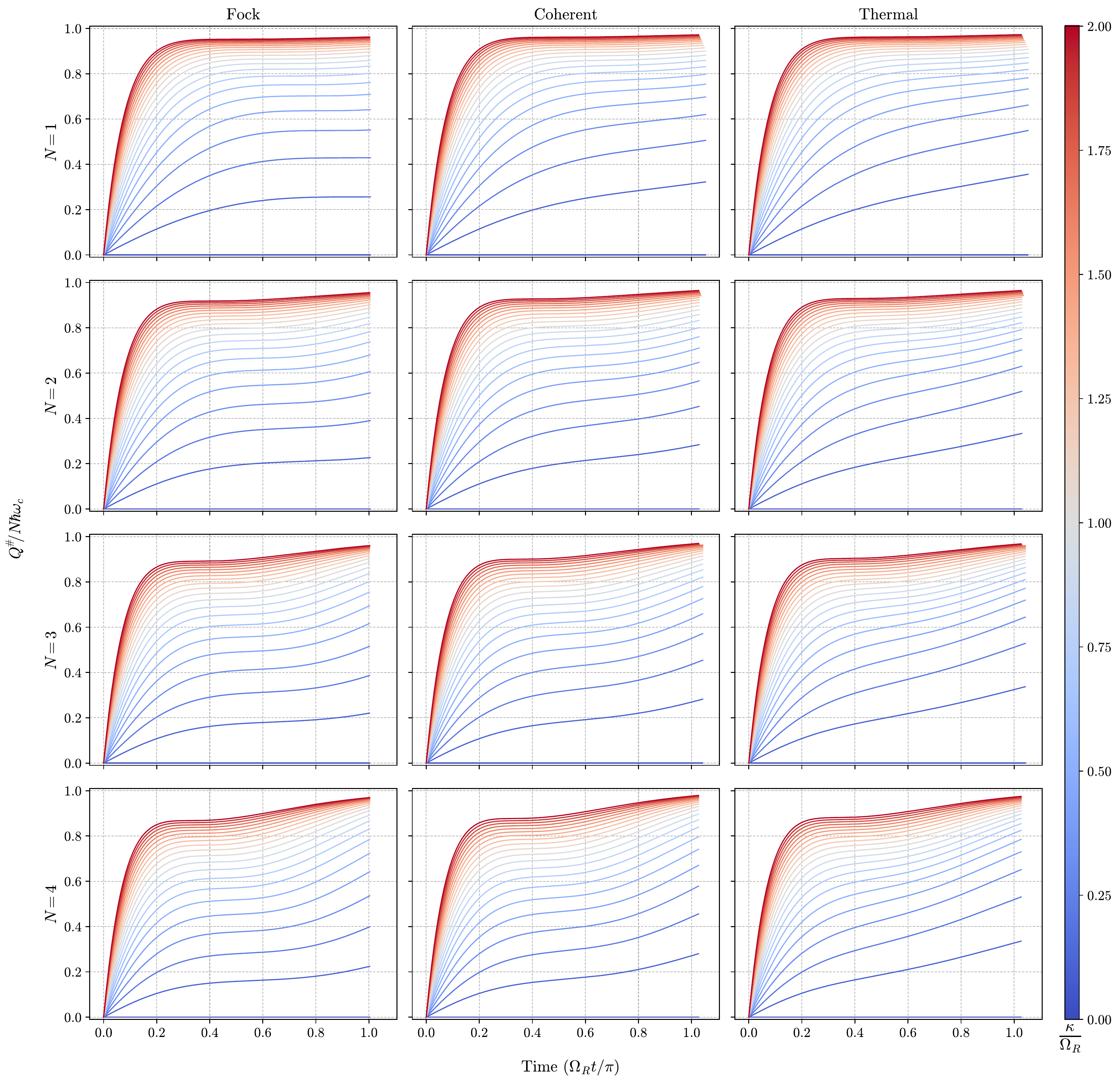}
    \caption{Heat from the QB over time up to near the first holder energy maximum for Fock, coherent and thermal states as the charger initial condition. Results of the collective QB version. These plots are very similar between the three cases of initial condition, with small differences only appreciable for $\kappa\approx 0$.}
    \label{fig:SM_fig8}
\end{figure*}

Simulation results complementary to Fig. \ref{fig:exampleRabiOsc+MaxPoint} of the main text are shown in Figs. \ref{fig:SM_fig5} to \ref{fig:SM_fig8}, showing the Rabi-like oscillations up to near the first energy maximum, for $N=1,..., 4$ and $\kappa \in [0, 2\Omega_R]$, for the three cases of charger initial conditions studied. These results correspond to the collective QB version.

The performances of the figures of merit are presented in Fig. \ref{fig:SM_fig1}. These results are very similar if we compare them with each other by fixing the figure of merit and changing the charger initial condition. The collective enhancements are presented in Fig. \ref{fig:SM_fig3}. These results are also very similar if we compare them with each other by fixing the figure of merit and changing the charger initial condition, excepting the ergotropy, as the appearance of zeros in the denominator appear in the Fock and Thermal cases, resulting in non-plotted data. Hence, the ergotropy is better analyzed with Fig. \ref{fig:SM_fig6}, which clearly shows that the coherent case is the more robust against dissipation, and explains the ergotropy shown in Fig. \ref{fig:result2} of the main text.

Further data and analyses can be found in \cite{Carrasco2021}, where the holder populations are plotted and a deeper study on the feasibility of using NV centers as qubits for Tavis-Cummings QBs is also done.


\providecommand{\noopsort}[1]{}\providecommand{\singleletter}[1]{#1}%
%


\end{document}